\documentstyle[12pt]{article}
\let\section=\subsection     \let\subsection=\subsubsection                %%
\begin{document}
\begin{center}
   {\large \bf PARTICLE INTERFEROMETRY}\\[2mm]
   {\large \bf IN HEAVY-ION COLLISIONS}\\[5mm]
   ULRICH HEINZ \\[5mm]
   {\small \it Institut f\"ur Theoretische Physik, Universit\"at Regensburg, \\
   D-93040 Regensburg, Germany \\[8mm] }
\end{center}

\begin{abstract}\noindent
 By measuring hadronic single-particle spectra and two-particle 
 correlations in heavy-ion collisions, the size and dynamical state of 
 the collision fireball at freeze-out can be reconstructed. I discuss 
 the relevant theoretical methods and their limitations. By applying 
 the formalism to recent pion correlation data from Pb+Pb collisions at 
 CERN we demonstrate that the collision zone has undergone strong 
 transverse growth before freeze-out (by a factor 2-3 in each 
 direction), and that it expands both longitudinally and 
 transversally. From the thermal and flow energy density 
 at freeze-out the energy density at the onset of transverse
 expansion can be estimated from conservation laws. It comfortably 
 exceeds the critical value for the transition to color deconfined 
 matter.  
 \end{abstract}

%%%%%%%%%%%%%%%%%%%%%%%%%%%%%%%%%%%%%%%%%%%%%%%%%%%%%%%%%%%%%%%%%%%%%%
\section{Introduction}
%%%%%%%%%%%%%%%%%%%%%%%%%%%%%%%%%%%%%%%%%%%%%%%%%%%%%%%%%%%%%%%%%%%%%% 

In the last few years a large body of evidence has been accumulated 
that the hot and dense collision region in ultrarelativistic heavy ion 
collisions thermalizes and shows collective dynamical behaviour. This 
evidence is based on a comprehensive analysis of the hadronic single 
particle spectra. It was shown that all available data on hadron 
production in heavy ion collisions at the AGS and the SPS can be 
understood within a simple model which assumes locally thermalized 
momentum distributions at freeze-out, superimposed by collective 
hydrodynamical expansion in both the longitudinal and transverse 
directions \cite{LH89,Stach94}. The collective dynamical behaviour in 
the transverse direction is reflected by a characteristic dependence 
of the inverse slope parameters of the $m_\perp$-spectra (``effective 
temperatures'') at small $m_\perp$ on the hadron masses \cite{LH89}. 
New data from the Au+Au and Pb+Pb systems \cite{QM96} support 
this picture and show that the transverse collective dynamics is much 
more strongly exhibited in larger collision systems than in the 
smaller ones from the first rounds of experiments. The amount of 
transverse flow also appears to increase monotonically with collision 
energy from GSI/SIS to AGS energies, but may show signs of saturation 
at the even higher SPS energy \cite{QM96}.  

The extraction of flow velocities and thermal freeze-out temperatures 
from the measured single particle spectra relies heavily on model 
assumptions \cite{LH89}. There have been alternative suggestions to 
explain the observed features of the hadron spectra without invoking 
hydro-flow \cite{L96,EHX96}. The single-particle spectra are ambiguous 
because they contain no direct information on the space-time structure 
and the space-momentum correlations induced by collective flow. In 
terms of the phase-space density at freeze-out (``emission function'') 
$S(x,p)$ the single-particle spectrum is given by $E\, dN/d^3p = \int 
d^4x\, S(x,p)$; the space-time information in $S$ is completely washed 
out by integration. Thus, on the single-particle level, comprehensive 
model studies are required to show that a simple hydrodynamical model 
with only a few thermodynamic and collective parameters can fit all 
the data, and additional consistency checks are needed to show that 
the extracted fit parameter values lead to an internally consistent 
theoretical picture. The published literature abounds with examples 
demonstrating that without such consistency checks the theoretical 
ambiguity of the single particle spectra is nearly infinite.  

This is the point where Bose-Einstein correlations between the momenta 
of identical particle pairs provide crucial new input. They give 
direct access to the space-time structure of the source {\it and}
its collective dynamics. In spite of some remaining model 
dependence the set of possible model sources can thus be reduced 
dramatically. The two-particle correlation function $C(q,K)$ is 
usually well approximated by a Gaussian in the relative momentum $q$ 
whose width parameters are called ``HBT (Hanbury~Brown-Twiss) radii''.  
It was recently shown \cite{HB95}-\cite{CNH95} that these radius 
parameters measure certain combinations of the second central 
space-time moments of the source. In general they mix the spatial and 
temporal structure of the source in a nontrivial way \cite{CSH95}, and 
the remaining model dependence enters when trying to unfold these 
aspects.  

Collective dynamics of the source leads to a dependence of the HBT 
radii on the pair momentum $K$; this has been known for many years 
\cite{P84,MS88}, but was recently quantitatively reanalyzed, both 
analytically \cite{CSH95,CNH95,AS95,CL96} and numerically 
\cite{WSH96,HTWW96}. The velocity gradients associated with 
collective expansion lead to a dynamical decoupling of different 
source regions in the correlation function, and the HBT radii measure
the size of the resulting ``space-time regions of homogeneity'' of the 
source \cite{MS88,AS95} around the point of maximum emissivity for 
particles with the measured momentum $K$. The velocity gradients are 
smeared out by a thermal smearing factor arising from the random 
motion of the emitters around the fluid velocity \cite{CSH95}. Due to 
the exponential decrease of the Maxwell-Boltzmann distribution, this 
smearing factor shrinks with increasing transverse momentum $K_\perp$ 
of the pair, which is the basic reason for the $K_\perp$-dependence of 
the HBT radii. 

Unfortunately, other gradients in the source (for example spatial and 
temporal temperature gradients) can also generate a $K$-dependence of 
the HBT radii \cite{CSH95,CL96}. Furthermore, the pion spectra in 
particular are affected by resonance decay contributions, but only at 
small $K_\perp$. This may also affect the HBT radii in a 
$K_\perp$-dependent way \cite{Schlei,WH96}. The isolation of 
collective flow, in particular transverse flow, from the 
$K_\perp$-dependence of the HBT radii thus requires a careful study of 
these different effects.  

We studied this $K$-dependence of the HBT radii within a simple 
analytical model for a finite thermalized source which expands both 
longitudinally and transversally. For presentation I use the 
Yano-Koonin-Podgoretskii (YKP) parametrization of the correlator 
which, for sources with dominant longitudinal expansion, provides an 
optimal separation of the spatial and temporal aspects of the source 
\cite{CNH95,HTWW96}. The YKP radius parameters are independent of the 
longitudinal velocity of the observer frame. Furthermore, in all 
thermal models without transverse collective flow, they show perfect 
$M_\perp$-scaling (in the absence of resonance decay contributions). 
Only the transverse gradients induced by a non-zero transverse flow 
can break this $M_\perp$-scaling, causing an explicit dependence on 
the particle rest mass. This allows for a rather model-independent 
identification of transverse flow from accurate measurements of the 
YKP correlation radii for pions and kaons. High-quality data should 
also allow to control the effects from resonance decays.  

Due to space limitations, I will be extremely selective with equations,
figures and references. A comprehensive and didactical discussion 
of the formalism and a more extensive selection of numerical examples 
can be found in Ref.~\cite{He96} to which I refer the reader for 
more details.

%%%%%%%%%%%%%%%%%%%%%%%%%%%%%%%%%%%%%%%%%%%%%%%%%%%%%%%%%%%%%%%%%%
\section{The source model}\label{sec2}
%%%%%%%%%%%%%%%%%%%%%%%%%%%%%%%%%%%%%%%%%%%%%%%%%%%%%%%%%%%%%%%%%%

For our study we use the model of Ref.~\cite{CNH95} for an
expanding thermalized source: 
 \begin{equation}
 \label{3.15}
    S(x,K)\! =\! {M_\perp \cosh(\eta{-}Y) \over 8 \pi^4 \Delta \tau}
    \exp\!\! \left[- {K{\cdot}u(x) \over T(x)}
                       - {(\tau-\tau_0)^2 \over 2(\Delta \tau)^2}
                       - {r^2 \over 2 R^2} 
                       - {(\eta- \eta_0)^2 \over 2 (\Delta \eta)^2}
           \right]
 \end{equation}
Here $r^2 = x^2+y^2$, the spacetime rapidity $\eta = {1 \over 2} 
\ln[(t+z)/(t-z)]$, and the longitudinal proper time $\tau= \sqrt{t^2-
z^2}$ parametrize the spacetime coordinates $x^\mu$, with measure 
$d^4x = \tau\, d\tau\, d\eta\, r\, dr\, d\phi$. $Y = {1\over 2} 
\ln[(E_K+K_L)/(E_K-K_L)]$ and $M_\perp = \sqrt{m^2 + K_\perp^2}$ 
parametrize the longitudinal and transverse components of the pair 
momentum ${\vec K}$.  $T(x)$ is the freeze-out temperature, $\sqrt{2} 
R$ is the transverse geometric (Gaussian) radius of the source, 
$\tau_0$ its average freeze-out proper time, $\Delta \tau$ the mean 
proper time duration of particle emission, and $\Delta \eta$ 
parametrizes \cite{CSH95} the finite longitudinal extension of the 
source. The expansion flow velocity $u^\mu(x)$ is parametrized as 
 \begin{equation}
 \label{26}
   u^\mu(x) = \left( \cosh \eta \cosh \eta_t(r), \,
                     \sinh \eta_t(r)\, {\vec e}_r,  \,
                     \sinh \eta \cosh \eta_t(r) \right) ,
 \end{equation}
with a boost-invariant longitudinal flow rapidity $\eta_l = \eta$ 
($v_l = z/t$) and a linear transverse flow rapidity profile 
 \begin{equation}
 \label{27}
  \eta_t(r) = \eta_f \left( {r \over R} \right)\, .
 \end{equation} 
$\eta_f$ scales the strength of the transverse flow. The exponent of 
the Boltzmann factor in (\ref{3.15}) can then be written as
 \begin{equation}
 \label{26a}
  K\cdot u(x) = M_\perp \cosh(Y-\eta) \cosh\eta_t(r) - 
                {\vec K}_\perp{\cdot}{\vec e}_r \sinh\eta_t(r)\, .
 \end{equation}
For vanishing transverse flow ($\eta_f=0$) the source depends only 
on $M_\perp$, and remains azimuthally symmetric for all $K_\perp$.                                                 

%%%%%%%%%%%%%%%%%%%%%%%%%%%%%%%%%%%%%%%%%%%%%%%%%%%%%%%%%%%%%%%%%%
\section{HBT radius parameters}\label{sec3}
%%%%%%%%%%%%%%%%%%%%%%%%%%%%%%%%%%%%%%%%%%%%%%%%%%%%%%%%%%%%%%%%%%

From the source function (\ref{3.15}) the correlation function is 
calculated via the relation \cite{S73}
  \begin{equation}
     C({\vec q},{\vec K})
      \approx  1 + {\left\vert \int d^4x\, S(x,K)\,
         e^{iq{\cdot}x}\right\vert^2 \over
         \left\vert \int d^4x\, S(x,K)\right\vert^2 }
      =  1 + {\left\vert \int d^4x\, S(x,K)\,
         e^{i{\vec q}{\cdot}({\vec x} - {\vec \beta} t)}\right\vert^2 \over
         \left\vert \int d^4x\, S(x,K)\right\vert^2 }\, .
  \label{1}
  \end{equation}
Here $q = p_1 - p_2$, $K = (p_1 + p_2)/2$ 
are the relative and average 4-momenta of the boson pair. The quality of
the approximation in (\ref{1}) is discussed in \cite{CSH95}. 
Since $p_1,p_2$ are on-shell and thus $K\cdot q=0$, $q^0 = 
{\vec \beta}{\cdot}{\vec q}$ (with ${\vec \beta} = {\vec K}/K^0\approx 
{\vec K}/E_K$) is not an independent variable and can be eliminated
(second equality in (\ref{1})). Therefore the Fourier transform in 
(\ref{1}) cannot be inverted without a model for $S(x,K)$. This is 
the reason for the model dependence of the interpretation of HBT 
correlation data mentioned in the Introduction.

We use a cartesian coordinate system with the $z$-axis along the beam 
direction and the $x$-axis along $\vec K_\perp$. Then $\vec \beta =
(\beta_\perp, 0, \beta_l)$. We assume an azimuthally symmetric source 
(impact parameter $\approx 0$) and write $C(\vec q, \vec K)$ in the 
YKP parametrization \cite{CNH95,HTWW96}:
 \begin{equation}
 \label{18}
   C({\vec q},{\vec K}) = 1 +  
     \exp\biggl[ - R_\perp^2 q_\perp^2 
                 - R_\parallel^2 \left( q_l^2 - (q^0)^2 \right) 
                 - \left( R_0^2 + R_\parallel^2 \right)
                         \left(q\cdot U\right)^2
                \biggr] .
 \end{equation}
Here $q_\perp^2 = q_x^2 + q_y^2$, and $R_\perp$, $R_\parallel$, 
$R_0$, $U$ are four $K$-dependent parameter functions. $U(\vec K)$ 
is a 4-velocity with only a longitudinal spatial component: 
 \begin{equation}
 \label{19}
   U({\vec K}) = \gamma({\vec K}) \left(1, 0, 0, v({\vec K}) \right) ,
   \ \ {\rm with} \ \
   \gamma = {1\over \sqrt{1 - v^2}}\, .
 \end{equation}
Its value depends, of course, on the measurement frame. The ``Yano-Koonin 
velocity'' $v(\vec K)$ can be calculated in an arbitrary reference frame 
from the second central space-time moments of $S(x,K)$ (for 
explicit expressions see Ref.~\cite{HTWW96}). It is, to a good 
approximation, the longitudinal velocity of the fluid element from 
which most of the particles with momentum $\vec K$ are 
emitted \cite{CNH95,HTWW96}. For sources with boost-invariant 
longitudinal expansion velocity the YK-rapidity associated with 
$v(\vec K)$ is linearly related to the pair rapidity $Y$ 
\cite{HTWW96}.

The other three YKP parameters do not depend on the longitudinal 
velocity of the observer. (This distinguishes the YKP 
form (\ref{18}) from other parametrizations \cite{P84,HB95,CSH95}.) 
Their physical interpretation is easiest in terms of coordinates 
measured in the frame where $v({\vec K})$ vanishes. There they are 
given by \cite{CNH95} 
 \begin{eqnarray}   
   R_\perp^2({\vec K}) &=& \langle \tilde y^2 \rangle \, ,
 \label{20a} \\
   R_\parallel^2({\vec K}) &=& 
   \left\langle \left( \tilde z - (\beta_l/\beta_\perp) \tilde x
                \right)^2 \right \rangle   
     - (\beta_l/\beta_\perp)^2 \langle \tilde y^2 \rangle 
     \approx \langle \tilde z^2 \rangle \, ,
 \label{20b} \\
   R_0^2({\vec K}) &=& 
   \left\langle \left( \tilde t -  \tilde x/\beta_\perp
                \right)^2 \right \rangle 
    -  \langle \tilde y^2 \rangle/\beta_\perp^2 
    \approx \langle \tilde t^2 \rangle .
 \label{20c}
 \end{eqnarray}
Here $\langle f(x) \rangle \equiv \int d^4x\, f(x)\, S(x,K) / \int 
d^4x \, S(x,K)$ denotes the ($K$-dependent) average over the source
function $S(x,K)$, and $\tilde x \equiv x - \bar x({\vec K})$ etc., 
where $\bar x({\vec K})=\langle x\rangle$ is (approximately) 
the source point with the highest intensity at momentum $\vec K$. 
$R_\perp$, $R_\parallel$ and $R_0$ thus measure, approximately, the 
($K$-dependent) transverse, longitudinal and temporal regions of 
homogeneity of the source in the local comoving frame of the emitter. 
The approximation in (\ref{20b},\ref{20c}) consists of dropping terms 
which (for our model) vanish in the absence of transverse flow and 
were found in \cite{HTWW96} to be small even for finite transverse 
flow. Note that it leads to a complete separation of the
spatial and temporal aspects of the source. This separation is spoiled
by sources with $\langle \tilde x^2 \rangle \ne \langle \tilde x^2 \rangle$.
For our source this happens for non-zero transverse (in particular for 
large) transverse flow $\eta_f$, but for opaque sources where particle 
emission is surface dominated \cite{HV96} this may be true even without 
transverse flow.

Since in the absence of transverse flow the $\beta$-dependent terms in 
(\ref{20b}) and (\ref{20c}) vanish and the source itself depends only 
on $M_\perp$, all three YKP radius parameters then show perfect 
$M_\perp$-scaling. Plotted as functions of $M_\perp$, they coincide 
for pion and kaon pairs (see Fig.~5 in \cite{He96}). For non-zero 
transverse flow this $M_\perp$-scaling is broken by two effects: (1) 
The thermal exponent (\ref{26a}) receives an additional contribution 
proportional to $K_\perp = \sqrt{ M_\perp^2 - m^2}$. (2) The terms 
which were neglected in the second equalities of (\ref{20b},\ref{20c}) 
are non-zero, and they also depend on $\beta_\perp = K_\perp/E_K$. 
Both effects induce an explicit rest mass dependence and destroy the 
$M_\perp$-scaling of the YKP size parameters.  

%%%%%%%%%%%%%%%%%%%%%%%%%%%%%%%%%%%%%%%%%%%%%%%%%%%%%%%%%%%%%%
\section{$M_\perp$-dependence of YKP radii and collective flow}
\label{sec4}
%%%%%%%%%%%%%%%%%%%%%%%%%%%%%%%%%%%%%%%%%%%%%%%%%%%%%%%%%%%%%%

Collective expansion induces corelations between coordinates and 
momenta in the source, and these result in a dependence of the HBT 
parameters on the pair momentum $K$. At each point in the source the 
local velocity distribution is centered around the average fluid 
velocity; two points whose fluid elements move rapidly relative to 
each other are thus unlikely to contribute particles with small 
relative momenta. Essentially only such regions in the source 
contribute to the correlation function whose fluid elements move with 
velocities close to the velocity of the observed particle pair. If the 
source expands rapidly and features large velocity gradients, these 
contributing ``regions of homogeneity'' will be small. Their size will 
be inversely related to the velocity gradients, scaled by a ``thermal 
smearing factor'' $\sqrt{T/M_\perp}$ which characterizes the width of 
the Boltzmann distribution \cite{CSH95}.  

Thus, for expanding sources, the HBT radius parameters are generically 
decreasing functions of the transverse pair mass $M_\perp$. The slope 
0f this decrease grows with the expansion rate \cite{WSH96,HTWW96}. 
Longitudinal expansion affects mostly the longitudinal radius 
parameter $R_\parallel$ and the temporal parameter $R_0$ 
\cite{HTWW96}; the latter is a secondary effect since particles from 
different points are usually emitted at different times, and a 
decreasing longitudinal homogeneity length thus also leads to a 
reduced effective duration of particle emission. The transverse 
radius parameter $R_\perp$ is invariant under longitudinal boosts and 
thus not affected at all by longitudinal expansion. It begins to drop 
as a function of $M_\perp$, however, if the source expands in the 
transverse directions. The sensitivity of $R_\parallel$ and $R_0$ to 
transverse flow is much weaker \cite{HTWW96}. Transverse 
(longitudinal) flow thus affects mostly the transverse (longitudinal) 
regions of homogeneity. For a quantitative study see Fig.~6 in 
Ref.~\cite{He96}.  

Unfortunately, the observation of an $M_\perp$-dependence of $R_\perp$
by itself is not sufficient to prove the existence of radial transverse 
flow. It can also be created by other types of transverse gradients,
e.g. a transverse temperature gradient \cite{CSH95,CL96}. To exclude 
such a possibility one must check the $M_\perp$-scaling of the YKP 
radii, i.e. the independence of the functions $R_i(M_\perp)$ 
($i=\perp,\parallel,0$) of the particle rest mass (which is not broken 
by temperature gradients). Since different particle species are 
affected differently by resonance decays, such a check further 
requires the elimination of resonance effects.  

%%%%%%%%%%%%%%%%%%%%%%%%%%%%%%%%%%%%%%%%%%%%%%%%%%%%%%%%%%%%%%
\section{Resonance decays}
\label{sec5}
%%%%%%%%%%%%%%%%%%%%%%%%%%%%%%%%%%%%%%%%%%%%%%%%%%%%%%%%%%%%%%

Resonance decays contribute additional pions at low $M_\perp$; these 
pions originate from a larger region than the direct ones, due to 
resonance propagation before decay. They cause an 
$M_\perp$-dependent modification of the HBT radii.  

Quantitative studies \cite{Schlei,WH96} have shown that the resonances 
can be subdivided into three classes with different characteristic 
effects on the correlator: \\
(i) Short-lived resonances with lifetimes up to a few fm/$c$ do not 
propagate far outside the region of thermal emission and thus affect 
$R_\perp$ only marginally. They contribute to $R_0$ and $R_\parallel$
up to about 1 fm via their lifetime; $R_\parallel$ is larger if pion 
emission occurs later because for approximately boost-invariant expansion 
the longitudinal velocity gradient decreases as a function of time.\\ 
(ii) Long-lived resonances with lifetimes of more than several hundred fm/$c$ 
do not contribute to the measurable correlation and thus only reduce the
correlation strength (the intercept at $q=0$), without changing the shape 
of the corelator. The reason is that they propagate very far before 
decaying, thus simulating a very large source which contributes to the 
correlation signal only for unmeasurably small relative momenta.\\
(iii) There is only one resonance which does not fall in either of these 
two classes and can thus distort the form of the correlation 
function: the $\omega$ with its lifetime of 23.4 fm/$c$. It contributes
a second bump at small $q$ to the correlator, giving it a non-Gaussian 
shape and complicating \cite{WH96} the extraction of HBT radii by a 
Gaussian fit to the correlation function. At small $M_\perp$ up 
to 10\% of the pions can come from $\omega$ decays, and this fraction
doubles effectively in the correlator since the other pion can 
be a direct one; thus the effect is not always negligible.  

In a detailed model study \cite{WH96} we showed that resonance 
contributions can be identified through the non-Gaussian features in 
the correlator induced by the tails in the emission function resulting 
from resonance decays. To this end one computes the second and fourth 
order $q$-moments of the correlator \cite{WH96}. The second order 
moments define the HBT radii, while the kurtosis (the normalized
fourth order moments) provide a lowest order measure for the 
deviations from a Gaussian shape. We found \cite{WH96} that, at least 
for the model (\ref{3.15}), a positive kurtosis can always be 
associated with resonance decay contributions. Strong flow also 
generates a non-zero, but small and apparently always negative 
kurtosis (see Fig.~12 in \cite{WH96}). Any $M_\perp$-dependence of 
$R_\perp$ which is associated with a positive $M_\perp$-dependent 
kurtosis must therefore be regarded with suspicion; an 
$M_\perp$-dependence of $R_\perp$ with a vanishing or negative 
kurtosis, however, cannot be blamed on resonance decays.  

\vspace*{7cm}
\includegraphics{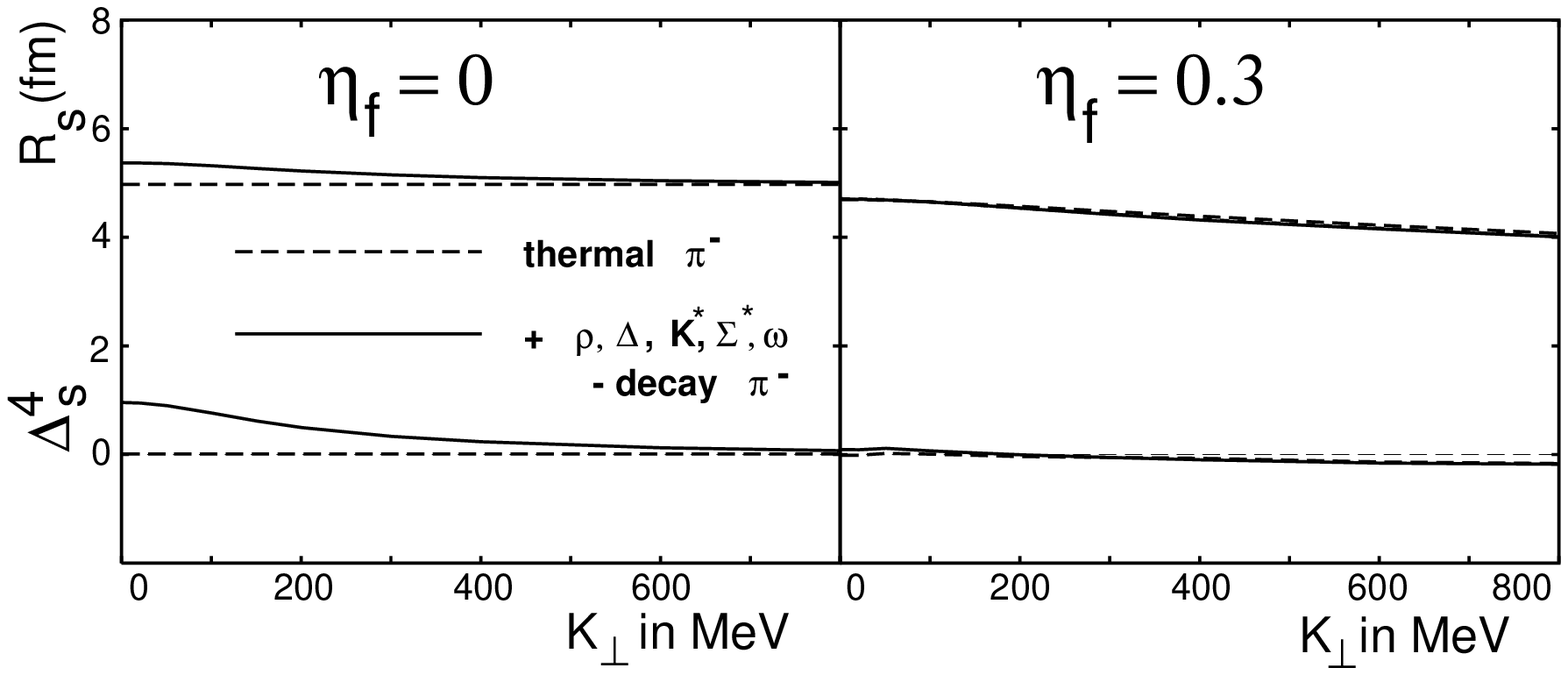}
\begin{center}
\begin{minipage}{13cm}
\baselineskip=12pt
{\begin{small}
{\bf Fig.~1.} 
The inverted $q$-variance $R_\perp$ and the kurtosis $\Delta_\perp$
(the index $s$ in the figure stands for ``sideward'') at $Y=0$ as
functions of $K_\perp$. Left: $\eta_f=0$ (no transverse flow). Right:
$\eta_f=0.3$. The difference between dashed and solid lines is
entirely dominated by $\omega$ decays.
\end{small}}
\end{minipage}
\end{center}

In our model, the first situation is realized for a source without 
transverse expansion (left panel of Fig.~1): At small $M_\perp$ 
the $\omega$ contribution increases $R_\perp$ by up to 0.5 fm while 
for $M_\perp > 600$ MeV it dies out. The effect on $R_\perp$ is small 
because the heavy $\omega$ moves slowly and doesn't travel very far 
before decaying. The resonance contribution is clearly visible in the 
positive kurtosis (lower curve). For non-zero transverse flow (right 
panel) there is no resonance contribution to $R_\perp$; this is 
because for finite flow the effective source size for the heavier 
$\omega$ is smaller than for the direct pions, and the $\omega$-decay 
pions thus always remain buried under the much more abundant direct 
ones. Correspondingly the kurtosis essentially vanishes; in fact, it 
is slightly negative, due to the weak non-Gaussian features induced by 
the transverse flow.  

%%%%%%%%%%%%%%%%%%%%%%%%%%%%%%%%%%%%%%%%%%%%%%%%%%%%%%%%%%%%%%%%%%%%
\section{Analysis of Pb+Pb data}
\label{sec6}
%%%%%%%%%%%%%%%%%%%%%%%%%%%%%%%%%%%%%%%%%%%%%%%%%%%%%%%%%%%%%%%%%%%%

In Fig.~2 we show a numerical fit of the YKP radius parameters, using
the expressions (\ref{20a})-(\ref{20c}) with our model source 
(\ref{3.15}), to data collected by the NA49 collaboration in 158 A 
GeV/$c$ Pb+Pb collisions \cite{NA49,Sch96}. Please note that this fit
refers to only a single rapidity slice of the available data, and it 
does not include resonance decays. The fit result must therefore be 
taken with great care. A comprehensive simultaneous analysis of all 
single particle spectra and two-particle correlation data from Pb+Pb 
collisions is in progress.  

\vspace*{13cm}
\includegraphics{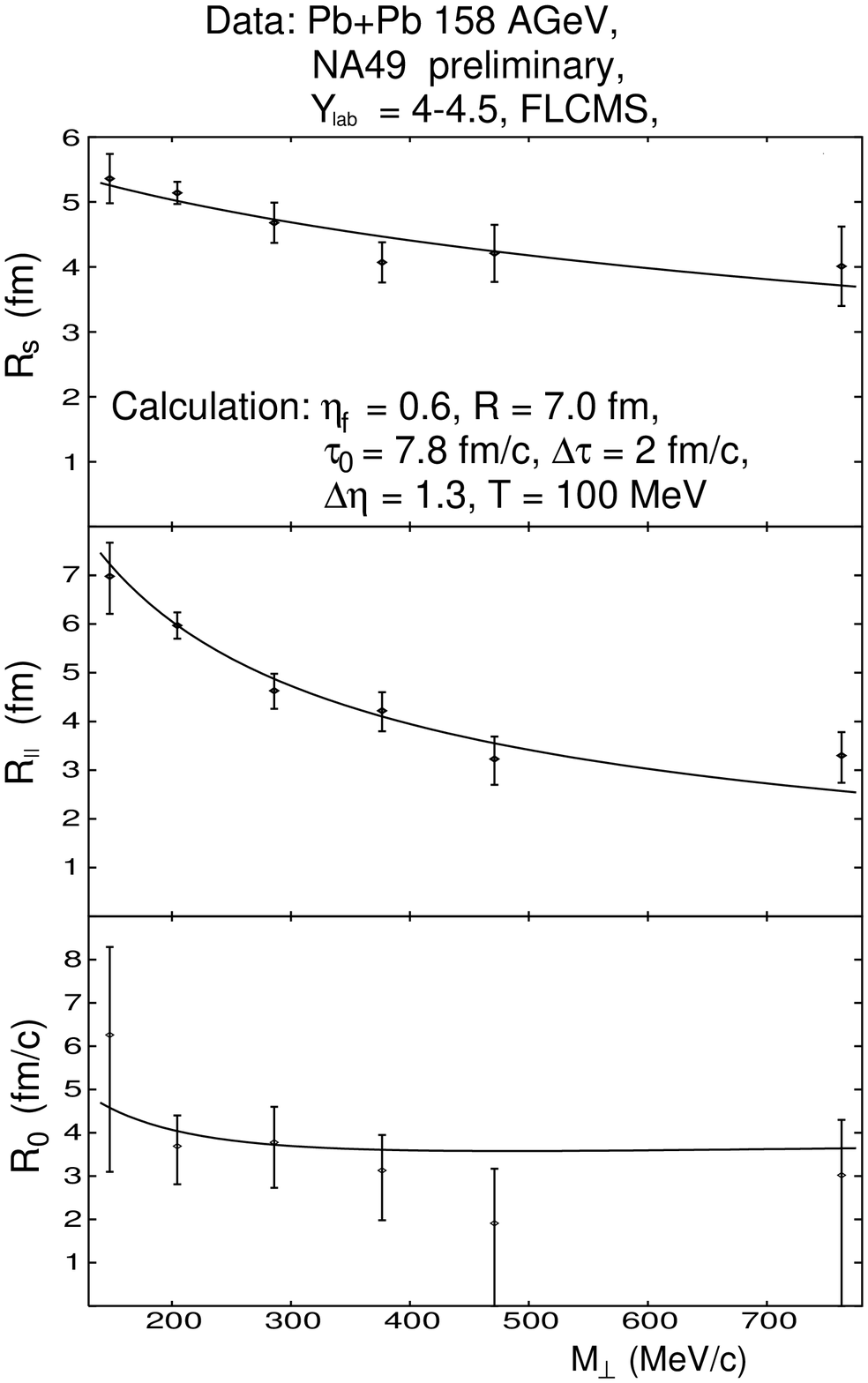}
\begin{center}
\begin{minipage}{13cm}
\baselineskip=12pt
{\begin{small}
{\bf Fig.~2.} 
 $R_\perp$, $R_\parallel$ and $R_0$ for 158 A GeV/$c$ Pb+Pb collisions 
 as functions of the transverse pair momentum. The data are from the 
 NA49 Collaboration \cite{Sch96}. The lines are a fit with the model 
 (\protect\ref{3.15}), with fit parameters as given in the figure.  
\end{small}} 
\end{minipage} 
\end{center} 

After $\Delta \eta{=}1.2$ has been adjusted to reproduce the 
width of the pion rapidity distribution \cite{Sch96}, the parameters 
$\tau_0$ and $\Delta \eta$ are essentially fixed by the magnitude of 
$R_\parallel$ and $R_0$. The radius $R$ is fixed by the magnitude of 
$R_\perp(K_\perp=0)$ once the temperature $T$ and transverse flow 
$\eta_f$ are known. The $M_\perp$-dependence of $R_\perp$ fixes $T$ 
and $\eta_f$, albeit not independently: essentially only the 
combination $\eta_f \sqrt{M_\perp/T}$, i.e. the velocity gradient 
divided by the thermal smearing factor, can be extracted 
\cite{CNH95,Sch96}. This is similar to the single particle spectra 
whose $M_\perp$-slopes determine only an effective blushifted 
temperature, $T_{\rm eff} = T \sqrt{{1+ \bar v_f \over 1- \bar v_f }}$ 
\cite{LH89}.  The correlations between $T$ and $\eta_f$ are, however, 
exactly opposite in the two cases: for a fixed spectral slope $T$ must 
be decreased if $\eta_f$ increases while a fixed $M_\perp$-slope of 
$R_\perp$ requires decreasing values of $\eta_f$ if $T$ is reduced 
\cite{Sch96}. The combination of single-particle spectra and 
two-particle correlation thus allows for a separate determination of 
$T$ and $\eta_f$.  

For the fit in Fig.~2 the freeze-out temperature was set by hand to 
$T=100$ MeV. The resulting flow parameter $\eta_f{=}0.6$ corresponds 
to an average transverse flow velocity $\bar v_f{=}0.58$. 
This combination of $T$ and $\eta_f$ results in single-particle 
spectra with roughly the right shape. Fitting $R_\perp$ with
higher temperatures results in larger $\eta_f$-values which leads to
single particle spectra which are much too flat.

Let us discuss in more detail the numbers resulting from this fit. 
First, the transverse size parameter $R{=}7$ fm is surprisingly large. 
Resonance contributions are not expected to reduce it by more than 0.5 
fm \cite{WH96}. The transverse flow correction to $R_\perp$ is 
appreciable, resulting in a visible transverse homogeneity length of 
only about 5.5 fm at small $K_\perp$, but even this number is large. 
$R{=}7$ fm corresponds to an r.m.s. radius $r_{\rm rms} = 
\sqrt{\langle \tilde x^2 + \tilde y^2 \rangle} \approx 10$ fm of the 
pion source, to be compared with an r.m.s. radius $r_{\rm rms}^{\rm 
Pb} = 1.2 \times A^{1/3} / \sqrt{5}$ fm = 3.2 fm for the density 
distribution of the original Pb nucleus projected on the transverse 
plane. This implies a transverse expansion of the reaction zone by a 
linear factor 3. That we also find a large transverse flow velocity 
renders the picture consistent. The longitudinal size of the collision 
region at the point where the pressure in the system began to drive 
the transverse expansion is less clear; assuming a similar expansion 
factor 3 in the beam direction (although we know from the fit that the 
longitudinal expansion velocity is much bigger than the transverse 
one) we conclude that the fireball volume must have expanded by at 
least a factor $3^3$ = 27 between the onset of transverse expansion 
and freeze-out! This is the clearest evidence for strong collective 
dynamical behaviour in ultra-relativistic heavy-ion collisions so far.  

The local comoving energy density at freeze-out can be estimated from 
the fitted values for $T$ and $\eta_f$. The thermal energy density of 
a hadron resonance gas at $T=100$ MeV and moderate baryon chemical 
potential is of the order of 50 MeV/fm$^3$. The large average 
transverse flow velocity of $\langle v_f \rangle \approx 0.58$ implies 
that about 50\% flow energy must be added in the lab frame. This 
results in an estimate of about $0.050$ GeV/fm$^3 \times 1.5 \times 27 
\approx 2$ GeV/fm$^3$ for the energy density of the reaction zone at 
the onset of transverse expansion.  This is well above the critical 
energy density $\epsilon_{\rm cr} \leq 0.9$ GeV/fm$^3$ predicted by 
lattice QCD for deconfined quark-gluon matter \cite{K96}. Whether this 
energy density was thermalized is, of course, a different question. In 
any case it must have been accompanied by transverse pressure, because 
otherwise transverse expansion could not have been initiated.  
                                        
%%%%%%%%%%%%%%%%%%%%%%%%%%%%%%%%%%%%%%%%%%%%%%%%%%%%%%%%%%%%%%%%%%%%
\section{Conclusions}
\label{sec7}
%%%%%%%%%%%%%%%%%%%%%%%%%%%%%%%%%%%%%%%%%%%%%%%%%%%%%%%%%%%%%%%%%%%%

I hope to have shown that

\begin{itemize}

\item
 two-particle correlation functions from heavy-ion collisions provide 
valuable information both on the geometry {\bf and} the dynamical 
state of the reaction zone at freeze-out;

\item
 a comprehensive and simultaneous analysis of single-particle spectra 
and two-particle correlations, with the help of models which provide
a realistic parametrization of the emission function, allows for an 
essentially complete reconstruction of the final state of the reaction 
zone, which can serve as a reliable basis for theoretical 
back-extrapolations towards the interesting hot and dense early stages 
of the collision; 

\item
 simple and conservative estimates, based on the crucial new 
information from HBT measurements on the large transverse size of the 
source at freeze-out and using only energy conservation, lead to the 
conclusion that in Pb+Pb collisions at CERN, before the onset of 
transverse expansion, the energy density exceeded comfortably the 
critical value for the formation of a color deconfined state of quarks 
and gluons.

\end{itemize}

\noindent
\begin{minipage}{14cm}
\baselineskip=11pt
{\footnotesize
 \noindent{\bf Acknowledgements:} 
 This work was supported by grants from DAAD, DFG, NSFC, BMBF and GSI. 
 The results reported here were obtained in collaboration with
 S. Chapman, P. Scotto, B. Tom\'a\v sik, U. Wiedemann, and Y.-F. Wu, to
 whom I would like express my thanks. I gratefully acknowledge many
 discussions with H. Appelsh\"auser, T. Cs\"org\H o, D. Ferenc, 
 M. Ga\'zdzicki, S. Sch\"onfelder, P. Seyboth, and A. Vischer.
}
\end{minipage} 

%%%%%%%%%%%%%%%%%%%%%%%%%%%%%%%%%%%%%%%%%%%%%%%%%%%%%%%%%%%%%%%%%%%%%%

\end{document}